\documentclass[12pt]{article}
\pdfoutput =1
\usepackage{graphicx}
\usepackage{float} 
\usepackage{subfloat}
\usepackage{amsmath}
\usepackage{slashed}
\usepackage{amsfonts}   
\usepackage{amssymb}
\usepackage[colorlinks]{hyperref}
\hypersetup{
  colorlinks,
  citecolor=blue,
  linkcolor=red,
  urlcolor=violet}
\usepackage{enumerate}
\numberwithin{equation}{section}
\usepackage{subcaption}
\usepackage{physics}
\usepackage{mathrsfs}

\begin{document}
\date{}
\title{{\bf{\Large JT gravity and the models of Hawking-Page transition for 2D black holes}}}
\author{
{\bf {\normalsize Arindam Lala}$
$\thanks{E-mail:  arindam.physics1@gmail.com, arindam.lala@pucv.cl}}\\
 {\normalsize  Instituto de F\'{i}sica, Pontificia Universidad Cat\'{o}lica de Valpara\'{i}so,}\\
  {\normalsize Casilla 4059, Valparaiso, Chile,}
  \\[0.3cm]
 {\bf {\normalsize Hemant Rathi\thanks{E-mail:  hrathi07@gmail.com, hrathi@ph.iitr.ac.in}~ and ~Dibakar Roychowdhury}$
$\thanks{E-mail:  dibakarphys@gmail.com, dibakar.roychowdhury@ph.iitr.ac.in}}\\
 {\normalsize  Department of Physics, Indian Institute of Technology Roorkee,}\\
  {\normalsize Roorkee 247667, Uttarakhand, India}
\\[0.3cm]
}
\maketitle
\begin{abstract}
We propose a top down construction for Jackiw-Teitelboim (JT) gravity using compactification of $ D=5 $ gravity theories in the presence of Abelian ($ U(1) $) as well as $ SU(2) $ Yang-Mills (YM) fields. The background solutions corresponding to $ D=2 $ model have been obtained in the perturbative regime of the theory where the corrections have been estimated over the uncharged JT solutions while treating the gauge couplings as the parameters of the expansion. Our analysis reveals the existence of two classes of solutions namely, (i) the \emph{interpolating} vacuum solution with AdS$_2$ in the IR and Lifshitz$_2$ in the UV (which serves as the thermal radiation background for our analysis) and (ii) the charged 2D black hole solution exhibiting Lifshitz$_2$ asymptotics. The analysis on thermal stability reveals the onset of a first order phase transition at $ T \sim T_0 $ such that for $ T < T_0 $ the only possible state is the thermal radiation background without any black hole. On the other hand, as the temperature is gradually increased beyond certain \emph{crossover} value $ T\sim T_2 (>T_0 )$, an emerging \emph{globally} stable black hole dominated phase has been observed which clearly indicates the onset of Hawking-Page (HP) transition in 2D gravity models.
\end{abstract}
\section{Overview and Motivation}
In today's literature, the celebrated AdS/CFT correspondence \cite{Maldacena:1997,Witten:1998ggc} is widely accepted to be the most remarkable achievement that took place in modern theoretical physics during last two decades. The underlying mechanism that stands behind this proposal is based on the so called holographic principle \cite{Susskind:1994vu} whose primary focus is to understand the strong coupling dynamics in large N QFTs using dual (weakly coupled) stringy/gravitational description living in higher dimensions. Since its discovery, several interesting examples have been discovered as well as tested within the realm of strong/weak conjecture among which the very recent discovery of an interesting toy model of quantum holography \cite{Kitaev:2014,Kitaev:2015} is something worthy of praise. This proposal is based on the original work of \cite{Sachdev:1993} and therefore goes under the name of Sachdev-Ye-Kitaev (SYK) models \cite{Polchinski:2016xgd,Maldacena:2016hyu}. 

Ever since this proposal has been put forward, there have been growing interests in order to explore the Large N near infra-red (IR) dynamics in SYK models using its dual gravitational description(s) that lives in one higher dimension. The collective analyses unveil $(1+1)$D Jackiw-Teitelboim (JT) dilaton gravity as the dual model \cite{Teitelboim:1983ux,Jackiw:1984je} that has been conjectured. Putting all these pieces together, finally leads towards an emergent SYK/AdS$ _2 $ correspondence \cite{Jevicki:2016bwu}-\cite{Roychowdhury:2018}. These analyses have been subsequently extended for charged SYK models and their corresponding dual gravitational counterparts \cite{Davison:2016ngz}-\cite{Lala:2019inz}.

The purpose of the present work is to elaborate an example of dimensional compactification which finally leads towards $(1+1)$D Jackiw-Teitelboim (JT) dilaton gravity in the presence of non-trivial matter (gauge) couplings. The parent 5D model \cite{Fan:2015aia,Fan:2015yza}, that we choose to start with, was actually proposed in order to construct electrically charged space-time solutions with anisotropic Lifshitz scaling \cite{Kachru:2008yh,Taylor:2008tg}. The Chern-Simons (CS) term in the original 5D model \cite{Fan:2015aia} was actually a supergravity-inspired ``$FFA$" type term \cite{Fan:2015aia, Romans:1985ps}. However, the dimensional reduction, on the other hand, results in some form of $ D=2 $ JT model where both gravity as well as the dilation are found to be non minimally coupled with the abelian (U(1)) and the $SU(2)$ Yang-Mills (YM) sectors of the theory. \\\\
Below we enumerate some notable facts about the $ D=2 $ model that is constructed in this paper.\\ \\ 
$ \bullet $ The ground state (vacuum) of the theory has been found to be \emph{interpolating} between a Lifshitz$_2$ at UV and an AdS$_2$ in the IR. This we identify as the thermal radiation background for our subsequent analysis in the Euclidean formalism.\\\\
$ \bullet $ Thermal excitations have been identified as \emph{charged} black holes with Lifshitz$_2$ asymptotics. In the Euclidean framework, these black holes serve as the basis for what we call the analogue of Hawking-Page (HP) transition \cite{Hawking:1983}-\cite{Myung:2007ti} in 2D gravity models.\\\\
$ \bullet $ Considering the Euclidean framework, we perform our analysis of thermal stability in black holes using a canonical ensemble. Our analysis reveals the existence of certain critical temperature, $ T \sim T_0 \sim \sqrt{\mu_0} $ such that for $ T> T_0 $ there exists two possible phases of black holes - one with lower mass and a negative heat capacity, and the other with higher mass and a positive heat capacity. As the temperature of the system is reduced below $ T_0 $, the black hole phase with negative heat capacity decays into pure thermal radiation via a \emph{first} order phase transition at $ T=T_0 $. On the other hand, as the temperature is increased above $ T_0 $, the lower mass black hole gradually transits into a globally stable phase of larger mass black hole that is in thermal equilibrium with its surrounding radiation background. During this course of transition, we observe several transition temperatures ($T_0< T_1< T_2$) and finally \emph{crossover} to a globally stable phase of black hole for $ T> T_2 $. We collectively identify all these features as being the analogue of HP transitions \cite{Hawking:1983} in ``charged" JT model whose holographic interpretation is yet to be unfolded.

The organisation for the rest of the paper is as follows: We construct our $ D=2 $ model in Sec. \ref{model} where we give a brief account for the dimensional reduction procedure and obtain the corresponding background solutions following a perturbative approach. Adopting to a Euclidean framework in Sec. \ref{phase}, we explore the thermal stability in black holes using a canonical ensemble. Finally, we conclude in Sec. \ref{conclusion}. 
\section{$D=2$ model}\label{model}
\subsection{A top down construction}
We consider $ D=5 $ gravity model minimally coupled to both abelian ($A_{M}$) as well as $SU(2)$ Yang-Mills (YM) fields ($A_{M}^{a}$ ($a=1,2,3$)) \cite{Fan:2015aia},
\begin{equation}\label{act:5D}
S_{5D}=\int d^{5}x \sqrt{-g}\left(\mathcal{R}-3\Lambda-\frac{\kappa}{4g_{s}^{2}}
F^{a}_{MN}F^{aMN}-\frac{\xi}{4}F_{MN}F^{MN}-\frac{\sigma}{2g_{2}^{2}}
\frac{\epsilon^{MNPQR}}{\sqrt{-g}}F^{a}_{MN}F^{a}_{PQ}A_{R}\right)
\end{equation}
where $ \Lambda (<0)$ is the cosmological constant\footnote{We set the AdS radius $L=1$ and $16\pi G=1$ in the subsequent analysis.}. Here  $\kappa$ and $\xi$ are the gauge coupling constants which would be treated as an expansion parameter in the subsequent analysis. The CS piece ($ \sim F^a \wedge F^a \wedge A $), on the other hand, is quite ubiquitous to $ D=5 $ theories and does not seem to have left with any of its imprints on the reduced $ D=2 $ model. 

The $ D=2 $ theory is obtained using the reduction ansatz \cite{Davison:2016ngz}
\begin{equation}\label{anz:red}
ds^{2}=\Phi^{\alpha}\;d\tilde{s}^{2}+\Phi^{\beta}dx_{i}^{2},\quad A_{M}^{a}dx^{M}
=A_{\mu}^{a}dx^{\mu}, \quad A_{M}dx^{M}=A_{\mu}dx^{\mu}~;~\alpha,\beta  \in \mathbb{R},
\end{equation}
together with the metric of the reduced spacetime,
\begin{equation}\label{met:2D}
d\tilde{s}^{2}=\tilde{g}_{\mu\nu}dx^{\mu}dx^{\nu}~; ~
g_{\mu\nu}=\Phi^{\alpha}\tilde{g}_{\mu\nu}~;~\mu , \nu= t, z.
\end{equation}

The $ D=2 $ action (modulo a total derivative) could be formally expressed as\footnote{The methodology of dimensional
reduction from $5D$ to $2D$ has been briefly discussed in Appendix \ref{dimred}.
}
\begin{eqnarray}\label{actsim:2D}
S_{2D}&=&\mathcal{V}_{3}\int d^{2}x\sqrt{-\tilde{g}}\;\Bigg(\tilde{\mathcal{R}}\Phi -V(\Phi)
-\Phi \mathcal{L}(\tilde{A}_{\mu},\tilde{A}^a_{\mu})\Bigg)+S_{\text{GH}}+S_{ct},       \nonumber\\
\mathcal{L}(\tilde{A}_{\mu},\tilde{A}^a_{\mu})&=&\frac{\kappa}{4g_{s}^{2}}
\tilde{F}^{a}_{\mu\nu}\tilde{F}^{a\mu\nu}+\frac{\xi}{4}\tilde{F}_{\mu\nu}\tilde{F}^{\mu\nu}~;
\quad V(\Phi)=3 \Lambda \Phi,
\end{eqnarray}
where, we set $\alpha=0$ and $\beta=2/3$ without any loss of generality. Notice that our model (\ref{actsim:2D}) is a special
case of \cite{Almheiri:2014cka} with $ C=0 $ and $ A=-3\Lambda $. Moreover, here $ S_{\text{GH}}=-\int dt \sqrt{-\gamma} 
\mathcal{K}\Phi $ is the standard Gibbons-Hawking (GH) term ($ \mathcal{K} $ being the extrinsic curvature associated with
the boundary hypersurface \cite{Gibbons:1976ue}) and $ S_{\text{ct}} $ is the so called counter term which cures the divergences of the on-shell
action near its asymptotic limits.
\subsection{Equations of motion}
The equations of motion that readily follow from the variation of the parent action (\ref{actsim:2D}) are listed below,
\begin{subequations}\label{eom:col}
\setlength{\jot}{8pt}
\begin{eqnarray}
\label{5a}
\left(\nabla_{\mu}\nabla_{\nu}-\tilde{g}_{\mu\nu}\Box\right)\Phi+\frac{\xi\Phi}{2}\left(
\tilde{F}_{\mu\rho}\tilde{F}_{\nu}^{\;\rho}-\frac{1}{4}\tilde{F}^{2}\tilde{g}_{\mu\nu}\right) \\\notag
\qquad\qquad\qquad\quad+\frac{\kappa\Phi}{2g_{s}^{2}}\left(\tilde{F}^{a}_{\mu\rho}
\tilde{F}_{\nu}^{a\rho}-\frac{1}{4}\tilde{F}^{a^2}\tilde{g}_{\mu\nu}\right)-\frac{3\Lambda}{2}
\Phi\tilde{g}_{\mu\nu}&=&0 \label{eom:guge}\\
\tilde{\mathcal{R}}-3\Lambda-\frac{\kappa}{4g_{s}^{2}}\tilde{F}^{a}_{\mu\nu}
\tilde{F}^{a\mu\nu}-\frac{\xi}{4}\tilde{F}_{\mu\nu}\tilde{F}^{\mu\nu}&=&0\label{eom:sclr}\\
\frac{1}{\sqrt{-\tilde{g}}}\partial_{\mu}\left(\sqrt{-\tilde{g}}\Phi\tilde{F}^{a\mu\nu}\right)
+\Phi\epsilon^{abc}\tilde{A}_{\mu}^{b}\tilde{F}^{c\mu\nu}&=&0\label{eom:nab}\\
\partial_{\mu}\left(\sqrt{-\tilde{g}}\Phi\tilde{F}^{\mu\nu}\right)&=&0. \label{eom:ab}
\end{eqnarray}
\end{subequations}

In order to proceed further, we choose to work with the static metric ansatz
\begin{equation}\label{met:ansz}
ds^{2}=e^{2\omega(z)}\left(-dt^{2}+dz^{2}\right)
\end{equation}
together with the ansatz for the gauge fields,
\begin{subequations}
\begin{align}
\tilde{A}_{\mu}&=\left(\tilde{A}_{t}(z),0\right)\label{ansz:ab}\\[10pt]
\tilde{A}_{\mu}^{a}&=\tilde{A}_{t}^{3}(z)\tau^{3}dt+\tilde{A}_{z}^{1}(z)\tau^{1}dz
\label{ansz:nab}
\end{align}
\end{subequations}
where $\tau^{a}=\sigma^{a}/2i$ are the Pauli matrices of $ SU(2) $ YM theory.

Using  (\ref{met:ansz}), (\ref{ansz:ab}) and (\ref{ansz:nab}) we finally note down the set of dynamical equations\footnote{In order to simplify our notations we set, $\tilde{A}_{t}^{3}=\chi(z)$ and $\tilde{A}_{z}^{1}=\eta(z)$.}, 
\begin{subequations}
\setlength{\jot}{8pt}
\begin{align}
2\Phi''+6\Lambda\Phi e^{2\omega}+\Phi e^{-2\omega}\left[\xi\left(\tilde{A}_{t}'\right)^{2}
+\frac{\kappa}{g_{s}^{2}}\left(\left(\chi'\right)^{2}+\chi^{2}\eta^{2}\right)\right]&=0 \label{nor:sclr1}\\
\Phi''-2\omega'\Phi'&=0 \label{nor:sclr2}\\
4\omega''+6\Lambda e^{2\omega}-e^{-2\omega}\left[\xi\left(\tilde{A}_{t}'\right)^{2}
+\frac{\kappa}{g_{s}^{2}}\left(\left(\chi'\right)^{2}+\chi^{2}\eta^{2}\right)\right]&=0\label{nor:guge}\\
\Phi \tilde{A}_{t}''+\Phi'\tilde{A}_{t}'-2\Phi\omega'\tilde{A}_{t}'&=0 \label{nor:ab}\\
\left(\Phi e^{-2\omega}\chi\eta\right)'+e^{-2\omega}\Phi\eta\chi'&=0 \label{nor:nab1}\\
\left(\Phi e^{-2\omega}\chi' \right)'-e^{-2\omega}\Phi\eta^{2}\chi&=0\label{nor:nab2}
\end{align}
\end{subequations}
together with the following constraint,
\begin{equation}\label{cnst:nab}
\Phi e^{-2\omega}\chi^{2}\eta =0.
\end{equation}
\subsection{Solving the dynamics}
We propose the following perturbative method of solving the dynamics\footnote{Here `$ab$' and `$na$' stand for respective perturbative corrections (to the uncharged JT solutions \cite{Almheiri:2014cka}) due to abelian and non-abelian sectors of the $ D=2 $ model (\ref{actsim:2D}).} (\ref{nor:sclr1})-(\ref{nor:nab2})
\begin{equation}\label{pert:exp}
\mathcal{A}(z)=\mathcal{A}_{(0)}(z)+\xi\mathcal{A}_{(1)}^{ab}(z)
+\kappa\mathcal{A}_{(1)}^{na}(z)+\cdots
\end{equation}
which is a perturbation in the gauge coupling parameters over the uncharged background solutions \cite{Almheiri:2014cka}. In the present analysis, we retain ourselves only upto leading order in the perturbative expansion (\ref{pert:exp}). The general strategy would be to substitute (\ref{pert:exp}) into (\ref{nor:sclr1})-(\ref{nor:nab2}) and obtain equations of motion at different order in the perturbative expansion\footnote{See Appendix \ref{apen:eom} for details.}. Here, $\mathcal{A}(z)$ stands for either of the dynamical variables $\Phi$, $\omega$, $\chi$ and $\eta$.\\ \\ 
Before we proceed further, the following observations are noteworthy to mention:\\ \\
$ \bullet $ Since we are interested in expanding the action (\ref{actsim:2D}) upto leading order in the gauge coupling, therefore it is sufficient to estimate leading/zeroth order solutions for both abelian and non abelian gauge fields.\\\\
$ \bullet $ The abelian sector (\ref{nor:ab}) could be further simplified in terms of other variables,
\begin{equation}
F_{zt}\equiv A_{t}'=\frac{Qe^{2\omega}}{\Phi}  \label{max:gen}
\end{equation}
where $Q$ is the corresponding $ U(1) $ charge. \\\\
$ \bullet $ The dilaton equation of motion (\ref{nor:sclr2}) could be recast as,
\begin{equation}\label{dilaton}
\Phi'(z)=-\frac{e^{2\omega}}{2}
\end{equation} 
whose solution may be expressed as,
\begin{equation}\label{dila:gen}
\Phi(z)\simeq -\int \frac{dz}{2}~  e^{2\omega}.
\end{equation}
\subsubsection{Interpolating vacuum}\label{sol:vac}
In order to find vacuum solutions, the first step would be to note down zeroth order solutions\footnote{See Appendix \ref{sol:zero} 
for the details of the derivation.} for both the dilaton and the metric from (\ref{dil:zero}) and (\ref{omga:zero}), 
 \begin{subequations}
\setlength{\jot}{5pt}
\begin{align}
e^{2\omega_{(0)}^{\text{vac}}}&=-\frac{2}{3\Lambda z^{2}}\label{met:vac}\\
\Phi_{(0)}^{\text{vac}}&=-\frac{1}{3\Lambda z}. \label{diln:vac}
\end{align}
\end{subequations}
Combining (\ref{met:vac}) and (\ref{met:ansz}) it is by now quite evident that the vacuum solution corresponds to AdS$_2$
spacetime in standard Poincare coordinates
\begin{equation}
ds^{2}\sim z^{-2}\qty(-dt{^2}+dz{^2}),
\end{equation}
where we identify $z\to 0$ as the asymptotic UV limit of the bulk space-time. On the other hand, $z\to\infty$ stands for the IR limit 
\cite{Almheiri:2014cka}.

Next, we note down solutions corresponding to (\ref{zero:nab1}) and (\ref{zero:nab2}) 
\begin{align}
\chi_{(0)}^{\text{vac}}&\simeq \log z \label{zero:chi}\\
\eta_{(0)}^{\text{vac}}&\simeq \frac{1}{z\left(1+\log 
z\right)^{2}} \label{zero:eta}
\end{align}
where we have multiplied (\ref{zero:nab2}) by $\chi_{(0)}/\eta_{(0)}$ and used (\ref{cnst:0}). 

On the other hand, using (\ref{omga:xi}) and (\ref{omga:kapa}), the solutions corresponding to $\omega_{(1)}^{ab}$ and $\omega_{(1)}^{na}$ can
be found as,
\begin{subequations}
\setlength{\jot}{5pt}
\begin{align}
\left(\omega_{(1)}^{ab}\right)^{\text{vac}}&=10z^{2}+\frac{\mathsf{C}}{z}
+\frac{Q^{2}\Lambda}{6}z^{2}\left(1-3\log z\right),\label{sol:w1ab}\\
\left(\omega_{(1)}^{na}\right)^{\text{vac}}&=-\frac{3\Lambda}
{z}.\label{sol:w1na}
\end{align}
\end{subequations}

Finally, the leading order solutions for dilaton may be obtained from (\ref{dil:xi}) and (\ref{dil:kapa}) 
\begin{subequations}
\setlength{\jot}{5pt}
\begin{align}
\left(\Phi_{(1)}^{ab}\right)^{\text{vac}}&=\frac{-3\left(\mathsf{C} -20z^{3}
\right)+4Q^{2}\Lambda z^{3}\left(1-\frac{3}{4}\cdot\log z\right)}{9\Lambda
z^{2}}, \label{sol:p1ab}\\
\left(\Phi_{(1)}^{na}\right)^{\text{vac}}&\simeq\frac{1}{z^{2}}.
\label{sol:p1na}
\end{align}
\end{subequations}

Combining all these pieces together we find,
\begin{align}\label{met:vacu}
\begin{split}
ds^{2}_{\text{vac}}&=e^{2\omega^{\text{vac}}}\left(-dt^{2}+dz^{2}\right) \\[6pt]
&\approx e^{2\omega_{(0)}^{\text{vac}}}\left(1+2\xi\left(\omega_{(1)}^{ab}
\right)^{\text{vac}}+2\kappa\left(\omega_{(1)}^{na}\right)^{\text{vac}}\right)
\left(-dt^{2}+dz^{2}\right).
\end{split}
\end{align}

Given the above metric structure (\ref{met:vacu}), it is customary to explore various asymptotic limits associated to it. For example, near the IR ($z\rightarrow\infty$) region one finds,
\begin{align}\label{met:IR}
\begin{split}
e^{2\omega^{\text{vac}}}_{IR}
&\simeq -\frac{2}{3\Lambda z^{2}}-\frac{2}{9\Lambda}\left(60\xi+Q^{2}\xi
\Lambda\left(1-3\log z\right)\right)+\mathcal{O}\left(z^{-3}\right)
\end{split}
\end{align}
which clearly reveals an emerging AdS$_2$ geometry.

On the other hand, the UV ($z\rightarrow0$) limit of the metric reveals,
\begin{align}\label{met:UV}
\begin{split}
e^{2\omega^{\text{vac}}}_{UV}
&\simeq \frac{1}{z^{3}}\left(-\frac{4\xi\mathsf{C}}{3\Lambda}+4
\kappa\right)-\frac{2}{3\Lambda z^{2}}  \\
&\quad -\frac{2}{9\Lambda}\left(60\xi+Q^{2}\xi\Lambda\left(1-3\log z\right)
\right)+\mathcal{O}\left(z^{2}\right)
\end{split}
\end{align}
an emerging Lifshitz$ _2 $ geometry with dynamical exponent $z_{\text{dyn}}=\frac{3}{2}$. A careful analysis further reveals that for $\xi=\kappa=0$, the resulting geometry becomes AdS$_2$ in both the asymptotic limits. This confirms that gauge fields in the theory are actually responsible for the change in asymptotic (UV) structure as found earlier in \cite{Lala:2019inz,Taylor:2008tg}.
\subsubsection{ 2D black holes}
The zeroth-order/ uncharged background solutions $\omega_{(0)}$ and $\Phi_{(0)}$ may be expressed as,
\begin{subequations}
\setlength{\jot}{5pt}
\begin{align}
e^{2\omega_{(0)}^{\text{BH}}}&=-\frac{8\mu}{3\Lambda\sinh^{2}
\left(2\sqrt{\mu}z\right)}  \label{met:bh}\\
\Phi_{(0)}^{\text{BH}}&=-\frac{2\sqrt{\mu}}{3\Lambda}\coth\left(
2\sqrt{\mu}z\right). \label{diln:bh}
\end{align}
\end{subequations}

Below we note down first-order solutions corresponding to the metric and the dilaton. In order to simplify our analysis, we consider the following change of coordinate,
\begin{equation}\label{coor:new}
z=\frac{1}{2\sqrt{\mu}}\coth^{-1}\left(\frac{\rho}{\sqrt{\mu}}\right).
\end{equation}

Using (\ref{coor:new}), the solution corresponding to (\ref{omga:xi}) can be expressed as,
\begin{align}\label{omga1a:BH}
\begin{split}
\left(\omega_{(1)}^{ab}\right)^{\text{BH}}&\simeq\frac{\rho}{\sqrt{\mu}}
+\left[\frac{\rho}{2\sqrt{\mu}}\cdot\log\left(\frac{\sqrt{\mu}+\rho}{\sqrt{\mu}
-\rho}\right)-1\right]  \\
&\quad -\frac{3Q^{2}\Lambda}{16\mu^{3/2}}\left\{2\sqrt{\mu}\left(1+\log\rho\right)+\rho\log
\rho\cdot\log\left(\frac{\sqrt{\mu}-\rho}{\sqrt{\mu}+\rho}\right)\right.  \\
&\qquad\qquad\qquad\quad \left. +\rho\left[\text{Li}\left(2,\frac{\rho}{\sqrt{\mu}}\right)-
\text{Li}\left(2,\frac{-\rho}{\sqrt{\mu}}\right)\right] \right\}.
\end{split}
\end{align}

On the other hand, substituting (\ref{omga:kapa}) in (\ref{dil:kapa}) and using (\ref{coor:new}) we find,
\begin{align}\label{omga1na:BH}
\begin{split}
\left(\omega_{(1)}^{na}\right)^{\text{BH}}\simeq\frac{\sqrt{\mu}\rho
+\left(\mu-\rho^{2}\right)\tanh^{-1}\left(\frac{\rho}{\sqrt{\mu}}\right)}
{2(\mu-1)\mu^{3/2}}.
\end{split}
\end{align}

Finally, we note down zeroth order solutions for gauge fields that readily follow from (\ref{zero:nab1}) and (\ref{zero:nab2}),
\begin{subequations}
\setlength{\jot}{5pt}
\begin{align}
\chi_{(0)}^{\text{BH}}&\simeq1-\frac{1}{2\sqrt{\mu}}
\log\left(\frac{\rho}{\sqrt{\mu}}\right)  \label{zero:chibh}  \\
\eta_{(0)}^{\text{BH}}&\simeq\frac{\left(\rho^{2}-\mu\right)}{4\rho
\mu^{3/2}}\left(1-\frac{1}{\sqrt{2\mu}}\log\left(\frac{\rho}{\sqrt{\mu}}
\right)\right)^{-2}.    \label{zero:etabh}
\end{align}
\end{subequations}

Collecting all the pieces together, we note down the black hole metric as,
\begin{align}\label{met:BH}
\begin{split}
ds^{2}_{\text{BH}}\simeq -\frac{8}{3\Lambda}\left(\rho^{2}-\mu\right)\left(1+2\xi\left(\omega_{(1)}^{ab}\right)^{\text{bh}}
+2\kappa\left(\omega_{(1)}^{na}\right)^{\text{bh}}\right)\left(-dt^{2}+\frac{d\rho^{2}}{4\left(
\rho^{2}-\mu\right)^{2}}\right).
\end{split}
\end{align}
The black hole solution (\ref{met:BH}) exhibits a horizon at $ \rho = \sqrt{\mu} $. On the other hand, substituting, $ \rho \sim \frac{1}{\delta} $ and thereby taking $ \delta \rightarrow 0 $ limit reveals, $ ds^{2}_{\text{BH}}\Big|_{\rho \rightarrow \infty} \sim  \frac{1}{\delta^3}$. This suggests that the UV asymptotic of the space time approaches Lifshitz$_2$ geometry with dynamical critical exponent $z_{\text{dyn}}=\tfrac{3}{2}$.

The Hawking temperature of the black hole (\ref{met:BH}) can be found as,
\begin{equation}\label{temp:BH}
T_{H}=\frac{\sqrt{\mu}}{\pi}.
\end{equation}

Finally, we note down the Wald entropy \cite{Wald:1993nt} of the black hole,\footnote{Here by entropy we mean entropy
per unit volume of the transverse space namely, $S_{W}/\mathcal{V}_{3}$.}
\begin{align}\label{ent:wald}
\begin{split}
S_{W}&=4\pi\Phi\big|_{\rho=\sqrt{\mu}}\\
&=-\frac{1}{96\Lambda\sqrt{\mu}(|\mu-1|)}\Big[(|\mu-1|)\Big\{-33Q^{2}\Lambda
\xi+64\mu\left(1+0.89\xi\right)   \\
&\qquad\qquad\qquad\qquad\quad-15Q^{2}\xi\left(\log\mu\right)\Big\}-240
\kappa\Big]
\end{split}
\end{align}
which clearly shows a discontinuity at $ \mu = \mu_0 =1 $. As we probe into details on thermal stability, we identify the above discontinuity as a signature of first order transition from an unstable black hole phase to pure radiation.

In order to compare the $2D$ black hole entropy (\ref{ent:wald}) with the black hole entropy in $5D$ we uplift the $2D$
solution into $5D$,
\begin{equation}
ds_{5D}^{2}=ds_{2D}^{2}+\Phi^{2/3}dx_{i}^{2}, \qquad i=1,2,3.
\end{equation}

In a manner similar to the $2D$ theory, the Wald entropy corresponding to the $5D$ black hole can be written as 
\cite{Jacobson:1993vj,Brustein:2007jj}
\begin{align}\label{wald:5D}
\begin{split}
\widehat{S}_{W}&=-2\pi \int_{\Sigma}\qty(\frac{\delta\mathscr{L}_{5D}}{\delta\mathcal{R}_{abcd}})
\hat{\epsilon}_{ab}\hat{\epsilon}_{cd}\bar{\epsilon}    \\
&=-2\pi \int_{\rho = \sqrt{\mu},t=\text{const.}}\qty(\frac{\delta\mathscr{L}_{5D}}{\delta
\mathcal{R}_{\rho t \rho t}})\hat{\epsilon}_{\rho t}\hat{\epsilon}_{\rho t}\Phi
\prod_{i=1}^{3}dx_{i}    \\
&=\mathcal{V}_{3}\; 4\pi\Phi \big|_{\rho=\sqrt{\mu}},
\end{split}
\end{align}
where $\Sigma$ is the codimension-$2$ space-like bifurcation surface with binormal $\hat{\epsilon}_{ab}$ which is symmetric
in $a \leftrightarrow b$ and normalized as $\hat{\epsilon}^{ab}\hat{\epsilon}_{ab}=-2$, $\bar{\epsilon}$ is the induced volume
on $\Sigma$, and $\mathcal{R}_{abcd}$ is the Riemann curvature tensor in the theory. Also the bifurcation surface is at the
horizon $\rho = \sqrt{\mu}$, $t=$ constant. Notice that, in deriving the final form of the Wald entropy in (\ref{wald:5D}) we have
used the fact that the Lagrangian density $\mathscr{L}_{5D}$ corresponding to the $5D$ theory (\ref{act:5D}) contains term linear 
in $\mathcal{R}$ only. In addition, we have used the form of the metric given in the ansatz (\ref{anz:red}) explicitly. Hence from
(\ref{ent:wald}) and (\ref{wald:5D}) we observe that the Wald entropy (per unit volume of the transverse space ($\mathcal{V}_{3}$))
for the two theories indeed match.

\section{Thermal stability and HP transition}\label{phase}
We now move on to the Euclidean formalism \cite{Witten:1998} and explore thermal stability in black holes. To start with, the full Euclidean \textit{onshell} action is schematically expressed as,
\begin{equation}\label{act:sep}
-S^{\text{os}}_{2D}=S_{\text{grav}}^{\text{os}}+S^{\text{os}}_{\text{GH}}+S^{\text{os}}_{ct}
\end{equation}
where each of the above entities on the R.H.S. of (\ref{act:sep}) are estimated using background solutions found in the previous Section. This is accompanied by an analytic continuation of the Lorentzian time, $ t\rightarrow i \tau $ with a periodicity $ \beta $. In the present analysis, there exists two distinct periodicities corresponding to thermal radiation bath ($ \beta_{\text{TH}} $) and the black hole phase ($ \beta_{\text{BH}} $) respectively. The black hole periodicity $ \beta_{\text{BH}} $ is uniquely fixed by (\ref{temp:BH}). On the other hand, the former ($ \beta_{\text{TH}} $) could be fixed by demanding that the ratio,
\begin{align}\label{fix:beta}
\begin{split}
\frac{\beta_{\text{TH}}}{\beta_{\text{BH}}}\approx 2+\mathcal{O}\left(\frac{1}{\rho_{c}}\right)
\end{split}
\end{align}
becomes independent of the radial cut-off $\rho=\rho_{c}(\rightarrow\infty)$ at large distances \cite{Witten:1998}.

Using (\ref{fix:beta}), the \emph{effective} free energy of the configuration could be obtained using the background subtraction method \cite{Witten:1998}, 
\begin{align}\label{dif:sos}
\begin{split}
\Delta \mathcal{F}
&=\mathcal{F}_{\text{BH}}-\mathcal{F}_{\text{TH}}   \\
&=\frac{1}{18\Lambda(\mu-1)}\Big[(\mu-1)\left\{-7Q^{2}\Lambda
\xi+16\xi\left(-30+0.22\mu\right)\right\}+120\Lambda\kappa\Big].
\end{split}
\end{align}

Notice that this free energy is quite unique to our original choices of parameters $\alpha$ and $\beta$ in (\ref{anz:red}) and
therefore to the JT gravity model proposed in (\ref{actsim:2D}). In other words, the background subtraction method used in this
paper is also quite pertinent to the choices of parameters $\alpha$ and $\beta$.

Finally, using a canonical ensemble ($ \mathcal{Z}\sim e^{-\Delta S^{\text{os}}_{2D}} $), the energy of the black hole configuration may be estimated as,
\begin{align}\label{avg:eng}
\mathcal{M}_{\text{BH}}\equiv \langle\mathcal{E}\rangle=\frac{1}{450\Lambda(\mu-1)^{2}}\left( 3000\kappa\Lambda(3\mu-1)-\xi(\mu-1)^{2}\left(12000
+175Q^{2}\Lambda+88\mu\right)\right).
\end{align}

\begin{figure}[h!]
\begin{subfigure}{.5\textwidth}
  \centering
  \includegraphics[width=1\linewidth]{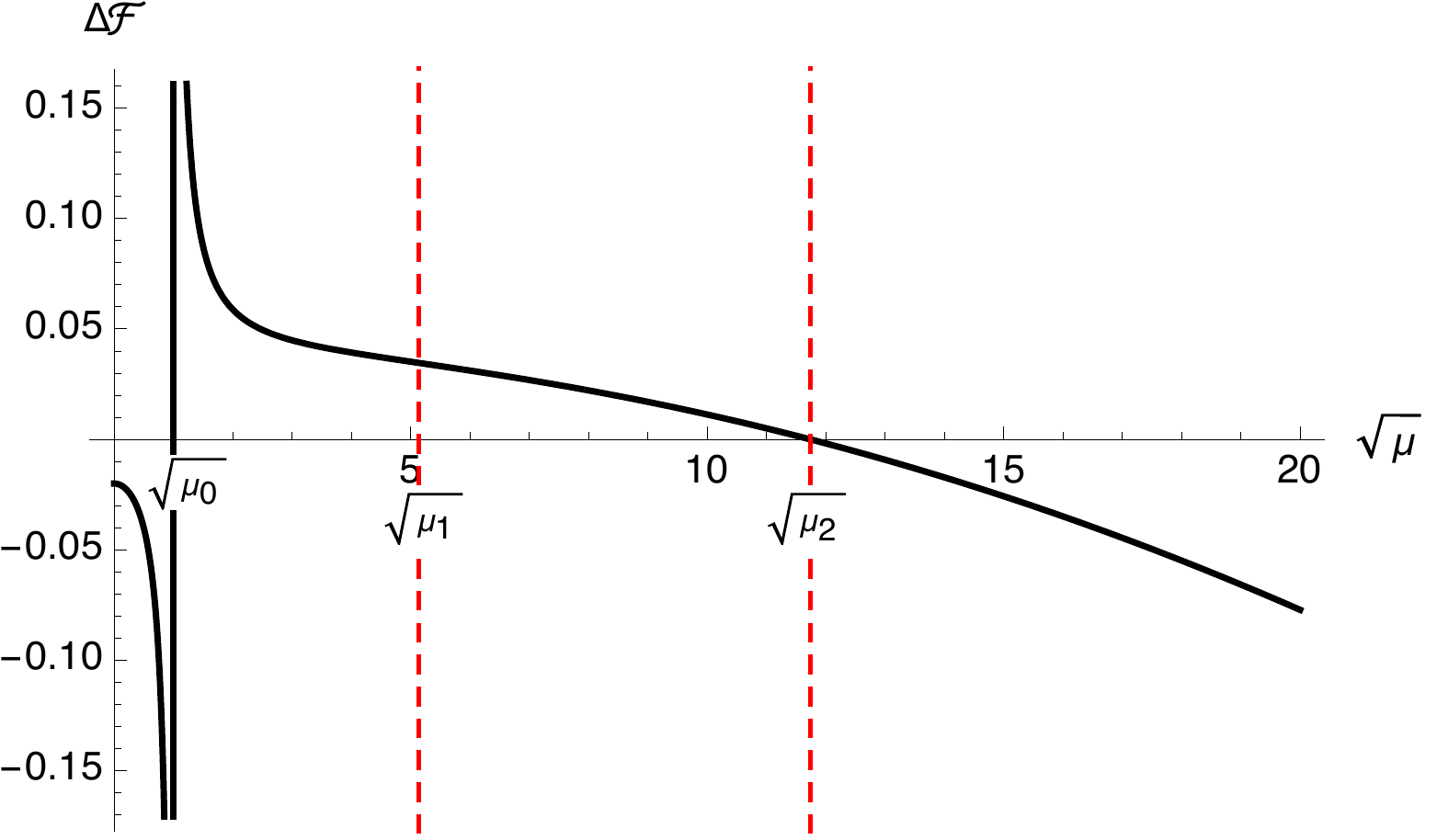} 
  \caption{Free energy \textit{vs.} temperature plot.} 
  \label{fig:1a}
\end{subfigure}
~~
\begin{subfigure}{.5\textwidth}
  \centering
  \includegraphics[width=1\linewidth]{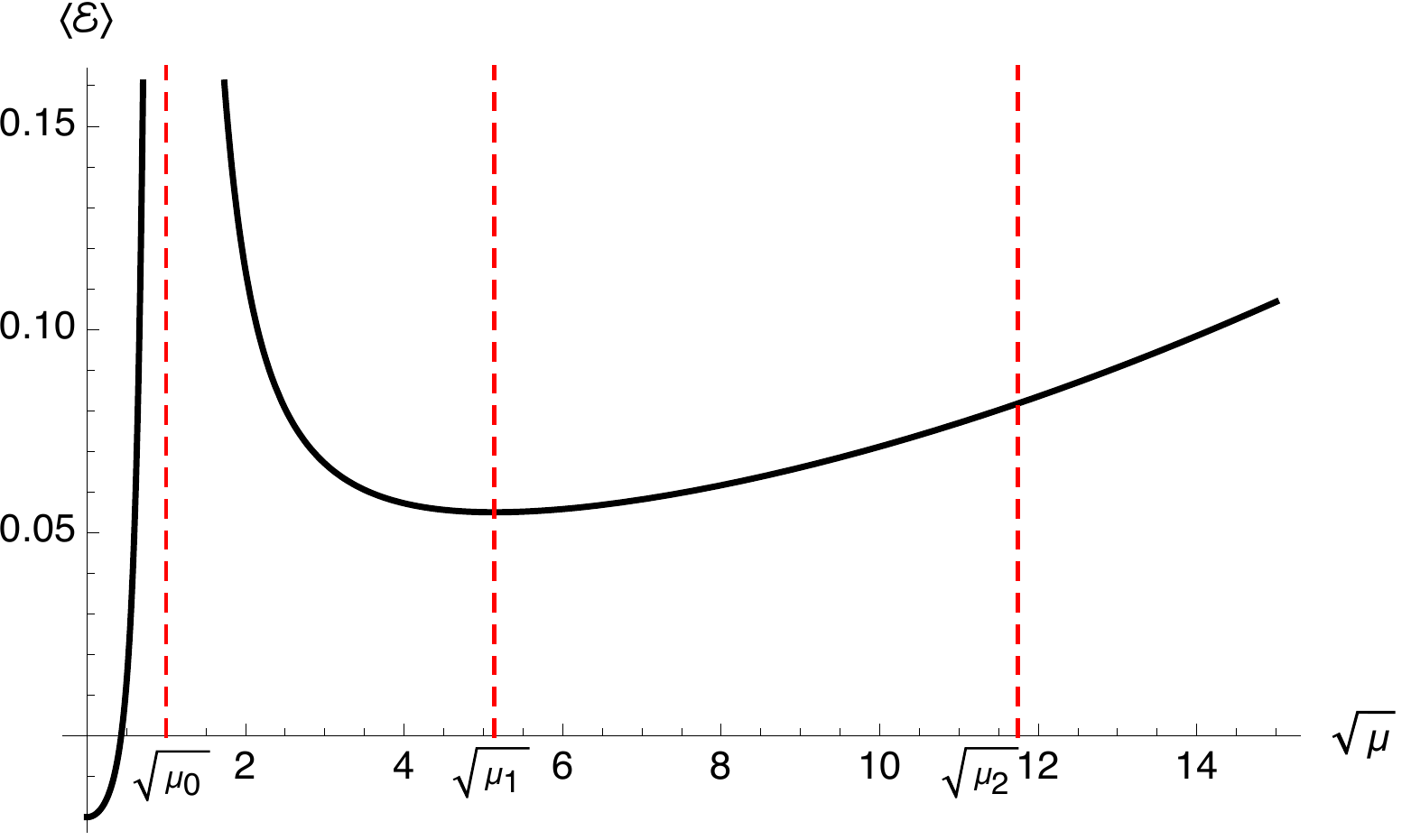}  
  \caption{Energy ($ \sim $Mass) \textit{vs.} temperature plot.}
  \label{fig:1b}
\end{subfigure}
\caption{Behaviour of effective free energy ($\Delta\mathcal{F}$) and
energy ($\mathcal{E} \sim \mathcal{M}_{\text{BH}}$) of the configuration with temperature
$T~ (\sim \sqrt{\mu})$.}
\label{fig:1}
\end{figure}

\begin{figure}[t!]
\begin{subfigure}{.5\textwidth}
  \centering
  \includegraphics[width=1\linewidth]{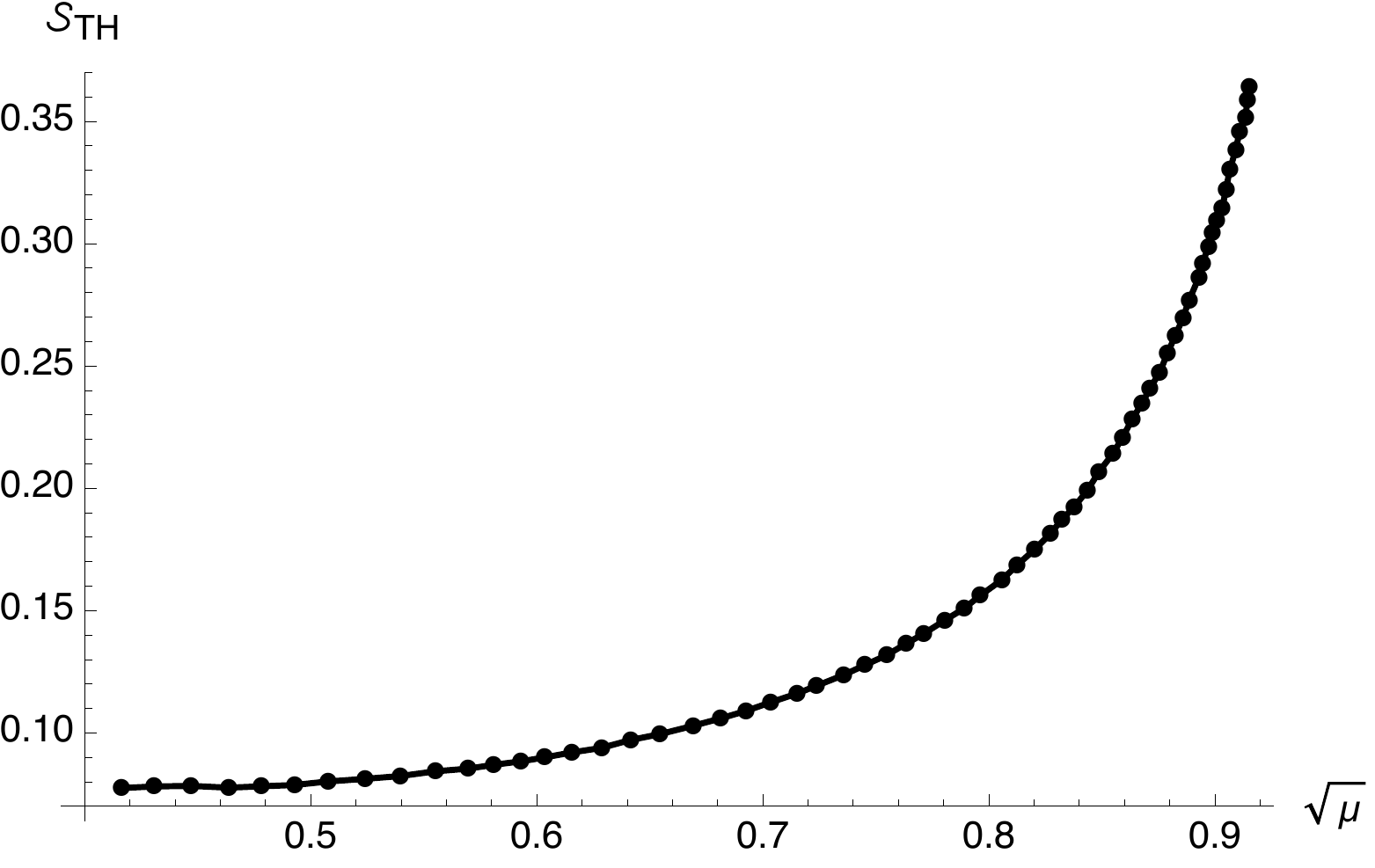}  
  \caption{Thermal entropy ($ \mathcal{S}_{\text{TH}} $) plot for $ \sqrt{\mu} <1 $ branch.}
  \label{fig:2a}
\end{subfigure}
~~
\begin{subfigure}{.5\textwidth}
  \centering
  \includegraphics[width=1\linewidth]{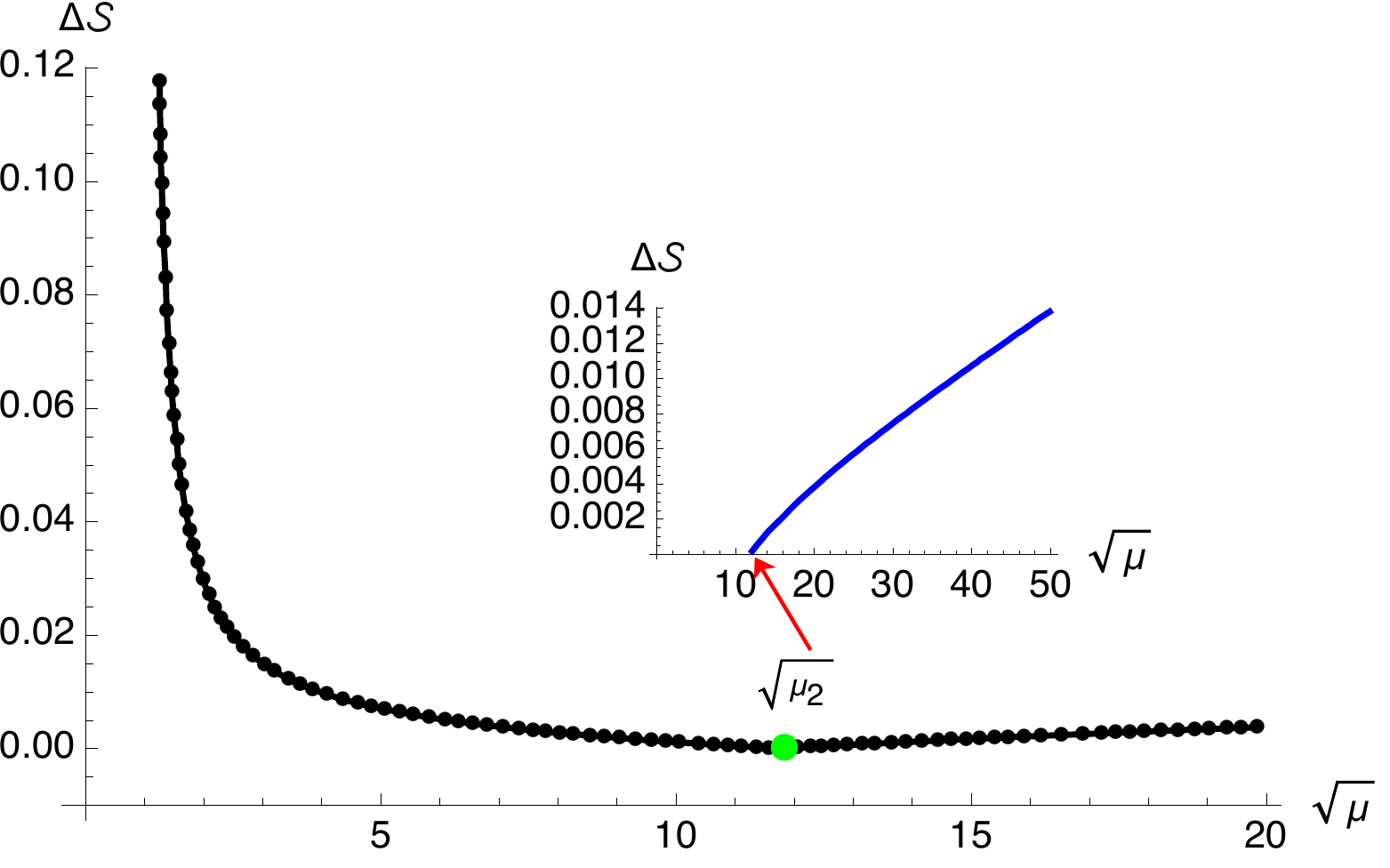}  
  \caption{Effective entropy ($ \Delta \mathcal{S} $) plot for the $\sqrt{\mu}>1$ branch.}
  \label{fig:2b}
\end{subfigure}
\caption{Behavior of the entropy ($ \mathcal{S} $) as a function of temperature $T(\sim
\sqrt{\mu})$.}
\label{fig:2}
\end{figure}

Fig.\ref{fig:1} displays the behavior of both the free energy ($ \Delta \mathcal{F} $) and the energy ($ \sim $ mass) of the (black hole) configuration with temperature ($ T \sim \sqrt{\mu} $). In all the subsequent plots, we set $Q=0.01$, $\xi=0.009$ and $\kappa=0.009$. Below we enumerate the key observations regarding Fig.\ref{fig:1} and Fig.\ref{fig:2}.\\\\
$ \bullet $ To start with, we notice that the thermal radiation collapses to form an \emph{unstable} black hole in the region $ T_0<T<T_1 $ which can either completely decay into pure radiation without a black hole (for $ T<T_0  $) or transit to a larger mass black hole for $ T>T_0 $. Clearly in this phase, the smaller mass black hole has larger free energy than the thermal radiation bath ($ \mathcal{F}_{\text{BH}}>\mathcal{F}_{\text{TH}} $) and is also characterized by a negative heat capacity, $ C_{\text{BH}} \sim \frac{\partial \langle\mathcal{E}\rangle}{\partial \sqrt{\mu}}<0 $ (Fig.\ref{fig:1b}). Therefore, we identify $ T_0<T<T_1 $ as a radiation dominated phase with $ \Delta \mathcal{F}>0 $ and $ T<T_0 $ as a pure radiation phase with, $ \Delta \mathcal{F}\sim \mathcal{F}_{\text{TH}}<0 $ (Fig.\ref{fig:1a}). The corresponding entropy\footnote{The entropy ($ \mathcal{S} $) that we measure should be understood in two ways. For $ \sqrt{\mu} <1$ branch, one is left with pure radiation, therefore we identify the corresponding entropy simply as the entropy of the thermal bath ($ \mathcal{S}_{\text{TH}} $). On the other hand, for $ \sqrt{\mu} >1$ branch, we introduce the notion of  \emph{effective} (or difference) entropy, $\Delta \mathcal{S} \sim -\frac{\Delta \mathcal{F}}{\Delta \sqrt{\mu}}\sim  |\mathcal{S}_{\text{BH}}-\mathcal{S}_{\text{TH}} |$ which is basically the difference in entropy between two distinct configurations namely, the black hole and the heat bath surrounding it (see (\ref{dif:sos})). However, in both cases, the change in total entropy must be positive following the second law of thermodynamics.} plot (Fig.\ref{fig:2}) reveals the onset of a first order phase transition at $T_0 \sim \sqrt{\mu_0} =1 $. The $ \sqrt{\mu}<1 $ branch (Fig.\ref{fig:2a}) corresponds to entropy of the configuration that ultimately boils down into pure radiation. This part of the phase diagram represents a rapidly evaporating black hole (Fig.\ref{fig:1b}) placed in a heat bath which eventually results in a thermal ensemble with \emph{unique} entropy ($ \mathcal{S}_{\text{TH}} $). The $\sqrt{\mu}>1$ branch (Fig.\ref{fig:2b}), on the other hand, depicts a phase with decreasing difference entropy ($\Delta \mathcal{S} $) for $ T\gtrsim T_0 $ which corresponds to the fact that the entropy of the unstable black hole approaches the entropy of the thermal bath as it gradually shrinks in size (Fig.\ref{fig:1b}) with the increase in temperature. This in turn is related to the negative heat capacity of the black hole as mentioned earlier.\\\\


$ \bullet $ As the temperature of the configuration approaches $ T \sim T_1 ~\sim \sqrt{\mu_1}$, the corresponding mass of the black hole reaches a minima (Fig.\ref{fig:1b} ) and thereafter starts increasing slowly for $ T\gtrsim T_1 $. We identify this as a \emph{locally}\footnote{By local we mean that this phase is existing only in a small temperature window $ T_1<T<T_2 $.} stable phase of black hole ($ \mathcal{S}_{\text{BH}} \sim \mathcal{S}_{\text{TH}} $) with positive heat capacity $ C_{\text{BH}} \sim \frac{\partial \langle\mathcal{E}\rangle}{\partial \sqrt{\mu}}>0 $ (Fig.\ref{fig:1b}). However, the effective free energy ($ \Delta \mathcal{F} $) of the configuration reveals that this phase is still dominated by the thermal radiation background with $ \mathcal{F}_{\text{TH}}<\mathcal{F}_{\text{BH}} $ (Fig.\ref{fig:1a}). The total entropy of the configuration is therefore $\sim \mathcal{S}_{\text{TH}}  $ and hence the difference $ \Delta \mathcal{S}\sim 0$ over the range $ T_1 <T<T_2 $ (Fig.\ref{fig:2b}).\\\\
$ \bullet $ Finally, we reach the point of \emph{crossover} $ T=T_2 $ beyond which the black hole mass ($ \mathcal{M}_{\text{BH}} $) starts becoming ever increasing with the increase in temperature and thereby yields a positive heat capacity ($ C_{\text{BH}} >0 $) (Fig.\ref{fig:1b}). A careful analysis further reveals that this is the point of inflection where the difference entropy vanishes ($ \Delta \mathcal{S}=0 $) and thereafter starts increasing (Fig.\ref{fig:2b}) as one approaches the region $ T>T_2 $. This corresponds to a \emph{globally}\footnote{By global we mean that this phase persists for all temperatures $ T>T_2 $.} stable phase of black hole with $ \mathcal{F}_{\text{BH}}<\mathcal{F}_{\text{TH}} $ (Fig.\ref{fig:1a}). Needless to say that for $ T>T_2 $, the entropy content of the configuration is mostly dominated due to the presence of the larger mass black hole which largely contributes to both in the increase in total entropy as well as the difference entropy ($ \Delta \mathcal{S} \gtrsim 0$) of the system.  
\section{Summary and final remarks}\label{conclusion}
We conclude with a brief summary of the main outcome in this analysis. We propose a possible formulation of $ D=2 $ Jackiw-Teitelboim (JT) gravity using dimensional reduction of parent $ D=5 $ gravity model \cite{Fan:2015aia} in the presence of abelian as well $ SU(2) $ Yang-Mills fields. To our surprise, we notice an early (thermal) instability associated to black hole micro-states using a canonical ensemble for the $ D=2 $ model. We identify that these early black hole states might either decay into pure radiation or they might switch over to a bigger black hole micro-states with higher entropy and lesser free energy. We summarise all these features collectively as being the analogue of Hawking-Page transition \cite{Hawking:1983} in 2D gravity models. This observation also rises a natural question regarding the holographic interpretation of our analysis in terms of (dual) quantum mechanical d.o.f. living in one dimension. We leave this as an exciting direction to be explored in the near future. 
\section*{Acknowledgments}
The work of A. L. is funded by the \emph{National Agency for Research and Development (ANID)}/ FONDECYT / POSTDOCTORADO BECAS CHILE / Project No. 3190021. H. R. and D. R. are indebted to the authorities of Indian Institute of Technology Roorkee for their unconditional support towards researches in basic sciences. 
\appendix

\section{Dimensional reduction from $5D$ to $2D$}\label{dimred}

In order to analyze the dimensional reduction of the $5D$ theory (\ref{act:5D}) to $2D$, in the following we rewrite the ansatz
given in (\ref{anz:red}):
\begin{equation}\label{anz:red2}
ds^{2}=\Phi^{\alpha}\;d\tilde{s}^{2}+\Phi^{\beta}dx_{i}^{2},\quad A_{M}^{a}dx^{M}
=A_{\mu}^{a}dx^{\mu}, \quad A_{M}dx^{M}=A_{\mu}dx^{\mu}~;~\alpha,\beta  \in \mathbb{R}.
\end{equation}

With this choice of the metric we note that
\begin{equation}\label{g:measure}
 \sqrt{-g}=\sqrt{-\tilde{g}}\;\Phi^{\alpha+3\beta/2},
\end{equation} 
where $\tilde{g}$ is the determinant of the two-dimensional metric.

It is then easy to check that the first two terms in the action (\ref{act:5D}) can be written as
\begin{align}\label{RL:5D}
\sqrt{-g}\qty(\mathcal{R}-3\Lambda)=\sqrt{-\tilde{g}}\;\Phi^{3\beta/2}\qty(\tilde{\mathcal{R}}
-3\Lambda\Phi^{\alpha}+\Phi^{\alpha}g^{ij}\mathcal{R}_{ij}). 
\end{align}

In the next step, we rewrite the third term within the curly braces in (\ref{RL:5D}) as a total derivative term given by
\begin{equation}\label{totDv}
\Phi^{\alpha+3\beta/2}g^{ij}\mathcal{R}_{ij}\sim \nabla_{\mu}\Big[g^{\mu\nu}
g^{\lambda\sigma}g_{\lambda\sigma}\left(\partial_{\nu}\Phi^{\beta}\right)
\Phi^{\alpha+\beta/2}\Big],
\end{equation}
which therefore does not contribute to the bulk equations of motion (\ref{5a})-(\ref{eom:ab}). Notice that, the JT gravity
model (\ref{actsim:2D}) can be obtained once we set the parameters $\alpha=0$ and $\beta=2/3$. In other words, these
choices of parameters are quite pertinent to the JT gravity model discussed in this paper. The remaining terms in
(\ref{actsim:2D}) involving gauge fields can be obtained in a similar manner.

\section{Equations of motion at different order in the perturbation series}\label{apen:eom}
In the following, we note down the equations of motion upto leading order in perturbative expansion. For example, substituting 
(\ref{pert:exp}) into (\ref{nor:sclr1}) we obtain
\begin{subequations}
\setlength{\jot}{8pt}
\begin{align}
\mathcal{O}(0): \quad& 2\Phi_{(0)}''+6\Lambda\Phi_{(0)} e^{2\omega_{(0)}}=0,  \label{dil:zero}\\
\mathcal{O}(\xi): \quad& 2\left(\Phi_{(1)}^{ab}\right)''+6\Lambda e^{2\omega_{(0)}}\left(2
\omega_{(1)}^{ab}\Phi_{(0)}+\Phi_{(1)}^{ab}\right)+\frac{Q^{2}}{\Phi_{(0)}} e^{2\omega_{(0)}}=0, 
\label{dil:xi}\\
\mathcal{O}(\kappa): \quad& 2\left(\Phi_{(1)}^{na}\right)''+6\Lambda e^{2\omega_{(0)}}\left(2 \omega_{(1)}^{na}\Phi_{(0)}+\Phi_{(1)}^{na}\right)+\frac{1}{g_{s}^{2}}\Phi_{(0)}e^{-2\omega_{(0)}}
\left(\left(\chi_{(0)}'\right)^{2}+\chi_{(0)}^{2}\eta_{(0)}^{2}\right)=0.  \label{dil:kapa}
\end{align}
\end{subequations}

Similarly (\ref{nor:guge}) could be arranged upto leading order in the perturbative expansion as
\begin{subequations}
\setlength{\jot}{8pt}
\begin{align}
\mathcal{O}(0): \quad& 4\omega_{(0)}''+6\Lambda e^{2\omega_{(0)}}=0,  \label{omga:zero}\\
\mathcal{O}(\xi): \quad& 4\left(\omega_{(1)}^{ab}\right)''+12\Lambda\omega_{(1)}^{ab} 
e^{2\omega_{(0)}}-\frac{Q^{2}}{\Phi_{(0)}^{2}}e^{2\omega_{(0)}}=0,  \label{omga:xi}\\
\mathcal{O}(\kappa): \quad& 4\left(\omega_{(1)}^{na}\right)''+12\Lambda\omega_{(1)}^{na} 
e^{2\omega_{(0)}}-\frac{e^{-2\omega_{(0)}}}{g_{s}^{2}}\left(\left(\chi_{(0)}'\right)^{2}
+\chi_{(0)}^{2}\eta_{(0)}^{2}\right)=0.  \label{omga:kapa}
\end{align}
\end{subequations}

Substituting (\ref{pert:exp}) into the constraint equation (\ref{cnst:nab}) we obtain
 \begin{subequations}
\setlength{\jot}{8pt}
\begin{align}
\mathcal{O}(0): &\quad \Phi_{(0)}e^{-2\omega_{(0)}}\chi_{(0)}^{2}\eta_{(0)}=0,  \label{cnst:0}\\
\mathcal{O}(\xi): &\quad  \Phi_{(0)}\left(\chi_{(0)}\eta_{(1)}^{ab}+2\eta_{(0)}\chi_{(1)}^{ab}\right)
+\chi_{(0)}\eta_{(0)}\left(\Phi_{(1)}^{ab}-2\Phi_{(0)}\omega_{(1)}^{ab}\right)=0, \qquad
e^{-2\omega_{(0)}}\chi_{(0)}^{2}=0, \label{cnst:xi}\\
\mathcal{O}(\kappa): &\quad\Phi_{(0)}\left(\chi_{(0)}\eta_{(1)}^{na}+2\eta_{(0)}\chi_{(1)}^{na}\right)
+\chi_{(0)}\eta_{(0)}\left(\Phi_{(1)}^{na}-2\Phi_{(0)}\omega_{(1)}^{na}\right)=0, \qquad
e^{-2\omega_{(0)}}\chi_{(0)}^{2}=0. \label{cnst:kapa}
\end{align}
\end{subequations}

Finally, from (\ref{nor:nab1}) and (\ref{nor:nab2}) the zeroth order equations could be recast as
\begin{subequations}
\setlength{\jot}{15pt}
\begin{align}
\partial_{z}\left(\Phi_{(0)}e^{-2\omega_{(0)}}\chi_{(0)}\eta_{(0)}\right)
+\Phi_{(0)}e^{-2\omega_{(0)}}\eta_{(0)}\left(\partial_{z}\chi_{(0)}\right)&=0,
\label{zero:nab1}\\
\partial_{z}\left(\Phi_{(0)}e^{-2\omega_{(0)}}\left(\partial_{z}\chi_{(0)}\right)
\right)-\Phi_{(0)}e^{-2\omega_{(0)}}\chi_{(0)}\eta_{(0)}^{2}&=0. \label{zero:nab2}
\end{align}
\end{subequations}

\section{Derivations of zeroth order solutions}\label{sol:zero}
In the conformal gauge \cite{Almheiri:2014cka}
\begin{equation}\label{conf:guge}
ds^{2}=-e^{2\omega(x^{+},x^{-})}dx^{+}dx^{-},\quad x^{\pm}=(t\pm z),
\end{equation}
the zeroth order equations of motion for the metric and the dilaton can be rewritten as
\begin{align}
4\partial_{+}\partial_{-}\Phi_{(0)}-3\Lambda\Phi_{(0)}e^{2\omega_{(0)}}&=0,\label{B1}\\ 
8\partial_{+}\partial_{-}\omega_{(0)}-3\Lambda e^{2\omega_{(0)}}&=0, \label{B2}\\
\partial_{+}\left(e^{2\omega_{(0)}}\partial_{+}\Phi_{(0)}\right)&=0, \label{B3}\\
\partial_{-}\left(e^{2\omega_{(0)}}\partial_{-}\Phi_{(0)}\right)&=0. \label{B4}
\end{align}

Clearly, the solution of (\ref{B2}) is given by
\begin{equation}\label{sol:B2}
e^{2\omega_{(0)}}=\left(-\frac{8}{3\Lambda}\right)\frac{1}{\left(x^{+}-x^{-}\right)^{2}}.
\end{equation}

Now integrating (\ref{B3}) and (\ref{B4}) we may write
\begin{align}
\partial_{+}\Phi_{(0)} &= \left(-\frac{8}{3\Lambda}\right)\frac{f(x^{-})}{\left(x^{+}
-x^{-}\right)^{2}},\label{sol:B3}\\
\partial_{-}\Phi_{(0)} &= \left(-\frac{8}{3\Lambda}\right)\frac{g(x^{+})}{\left(x^{+}
-x^{-}\right)^{2}}.\label{sol:B4}
\end{align}

Differentiating (\ref{sol:B3}) w.r.to $x^{-}$ and (\ref{sol:B4}) w.r.to $x^{+}$ and substituting them in (\ref{B1}) we obtain
\begin{align}\label{sol:phigen}
  \Phi_{(0)} =
    \begin{cases}
      \frac{2}{3\Lambda}\frac{\partial_{-}f(x^{-})\cdot (x^{+}-x^{-})+2f(x^{-})}
      {(x^{+}-x^{-})}  \\[10pt]
      \frac{2}{3\Lambda}\frac{\partial_{+}g(x^{+})\cdot (x^{+}-x^{-})-2g(x^{+})}
      {(x^{+}-x^{-})} .
    \end{cases}       
\end{align}

After a few easy algebraic steps, the general solution for the dilaton may be written as \cite{Almheiri:2014cka}
\begin{equation}\label{sol:phiF}
\Phi_{(0)}=\left(-\frac{2}{3\Lambda}\right)\frac{a+b(x^{+}+x^{-})+cx^{+}x^{-}}
{(x^{+}-x^{-})},
\end{equation}
where $a$, $b$, $c$ are real constants. For the vacuum we may choose $a=1$, $b=0$ and $c=0$ \cite{Almheiri:2014cka,
Yoshida1:2017}.

In the next step, we find the corresponding solutions for the black hole by exploiting the $SL(2,\mathbb{R})$ invariance of the 
metric (\ref{conf:guge}). The general solutions can then be written as
\begin{align}
e^{2\omega_{(0)}}&=\left(-\frac{8}{3\Lambda}\right)\frac{\omega^{+}(x^{+})
\omega^{-}(x^{-})}{\left[\omega^{+}(x^{+})-\omega^{-}(x^{-})\right]^{2}}, 
\label{met:w}\\[10pt]
\Phi_{(0)}&=\left(-\frac{2}{3\Lambda}\right)\frac{1-\mu\;\omega^{+}(x^{+})
\omega^{-}(x^{-})}{\left[\omega^{+}(x^{+})-\omega^{-}(x^{-})\right]}, \label{dil:w}
\end{align}
where $\mu(>0)$ can be considered as the mass of the black hole, and $\omega^{\pm}(x^{\pm})$ are \emph{monotonic}
functions \cite{Almheiri:2014cka}. 

Finally, using the conformal transformations
\begin{equation}
\omega^{\pm}(x^{\pm})=\frac{1}{\sqrt{\mu}}\tanh\sqrt{\mu}x^{\pm}
\end{equation}
(\ref{met:w}) and (\ref{dil:w}) can be rewritten as
\begin{align}
e^{2\omega_{(0)}}&=\left(-\frac{8}{3\Lambda}\right)\frac{\mu}{\sinh^{2}\left(
\sqrt{\mu}(x^{+}-x^{-})\right)}, 
\label{met:H}\\[10pt]
\Phi_{(0)}&=\left(-\frac{2}{3\Lambda}\right)\sqrt{\mu}\coth\left(\sqrt{\mu}(x^{+}
-x^{-})\right). \label{dil:H}
\end{align}



\begin{thebibliography}{99}

\bibitem{Maldacena:1997} J.~M.~Maldacena,
  ``The Large N limit of superconformal field theories and supergravity,''
  Int.\ J.\ Theor.\ Phys.\  {\bf 38} (1999) 1113
   [Adv.\ Theor.\ Math.\ Phys.\  {\bf 2} (1998) 231]
  doi:10.1023/A:1026654312961, 10.4310/ATMP.1998.v2.n2.a1
  [hep-th/9711200].

\bibitem{Witten:1998ggc} E.~Witten,
  ``Anti-de Sitter space and holography,''
  Adv.\ Theor.\ Math.\ Phys.\  {\bf 2} (1998) 253
  doi:10.4310/ATMP.1998.v2.n2.a2
  [hep-th/9802150].
  
  \bibitem{Susskind:1994vu}
L.~Susskind,
``The World as a hologram,''
J. Math. Phys. \textbf{36}, 6377-6396 (1995)
doi:10.1063/1.531249
[arXiv:hep-th/9409089 [hep-th]].
  
  
  \bibitem{Kitaev:2014}A. Kitaev. 2014. Hidden correlations in the Hawking radiation
  and thermal noise, talk given at Fundamental Physics Prize Symposium,
  November 10, Santa Barbara, U.S.A.
  
  \bibitem{Kitaev:2015}A. Kitaev. 2015. A simple model of quantum holography,
  talk given at KITP strings seminar and Entanglementprogram, February 12,
  April 7, and May 27, Santa Barbara, U.S.A.
  
   \bibitem{Sachdev:1993} S.~Sachdev and J.~Ye, Gapless spin fluid
  ground state in a random, quantum Heisenberg magnet, Phys. Rev.
  Lett. 70 (1993) 3339 [cond-mat/9212030]
  
  \bibitem{Polchinski:2016xgd} 
  J.~Polchinski and V.~Rosenhaus,
  ``The Spectrum in the Sachdev-Ye-Kitaev Model,''
  JHEP {\bf 1604}, 001 (2016)
  doi:10.1007/JHEP04(2016)001
  [arXiv:1601.06768 [hep-th]].
  
  \bibitem{Maldacena:2016hyu} 
  J.~Maldacena and D.~Stanford,
  ``Remarks on the Sachdev-Ye-Kitaev model,''
  Phys.\ Rev.\ D {\bf 94}, no. 10, 106002 (2016)
  doi:10.1103/PhysRevD.94.106002
  [arXiv:1604.07818 [hep-th]].
  
   \bibitem{Teitelboim:1983ux} 
  C.~Teitelboim,
  ``Gravitation and Hamiltonian Structure in Two Space-Time Dimensions,''
  Phys.\ Lett.\  {\bf 126B}, 41 (1983).
  doi:10.1016/0370-2693(83)90012-6
  
  \bibitem{Jackiw:1984je} 
  R.~Jackiw,
  ``Lower Dimensional Gravity,''
  Nucl.\ Phys.\ B {\bf 252}, 343 (1985).
  doi:10.1016/0550-3213(85)90448-1  
  
  \bibitem{Jevicki:2016bwu} 
  A.~Jevicki, K.~Suzuki and J.~Yoon,
  ``Bi-Local Holography in the SYK Model,''
  JHEP {\bf 1607}, 007 (2016)
  doi:10.1007/JHEP07(2016)007
  [arXiv:1603.06246 [hep-th]].
  
  \bibitem{Gross:2016kjj} 
  D.~J.~Gross and V.~Rosenhaus,
  ``A Generalization of Sachdev-Ye-Kitaev,''
  JHEP {\bf 1702}, 093 (2017)
  doi:10.1007/JHEP02(2017)093
  [arXiv:1610.01569 [hep-th]].
  
  \bibitem{Gross:2017hcz} 
  D.~J.~Gross and V.~Rosenhaus,
  ``The Bulk Dual of SYK: Cubic Couplings,''
  JHEP {\bf 1705}, 092 (2017)
  doi:10.1007/JHEP05(2017)092
  [arXiv:1702.08016 [hep-th]].
  
  \bibitem{Kitaev:2017awl} 
  A.~Kitaev and S.~J.~Suh,
  ``The soft mode in the Sachdev-Ye-Kitaev model and its gravity dual,''
  JHEP {\bf 1805}, 183 (2018)
  doi:10.1007/JHEP05(2018)183
  [arXiv:1711.08467 [hep-th]].
  
 \bibitem{Taylor:2017dly} 
  M.~Taylor,
  ``Generalized conformal structure, dilaton gravity and SYK,''
  JHEP {\bf 1801}, 010 (2018)
  doi:10.1007/JHEP01(2018)010
  [arXiv:1706.07812 [hep-th]].
  
  \bibitem{Moitra:2019}
  U.~Moitra, S.~K.~Sake, S.~P.~Trivedi and V.~Vishal,
  ``Jackiw-Teitelboim Model Coupled to Conformal Matter in the Semi-Classical Limit,''
  arXiv:1908.08523 [hep-th].

   \bibitem{Almheiri:2014cka} 
  A.~Almheiri and J.~Polchinski,
  ``Models of AdS$_{2}$ backreaction and holography,''
  JHEP {\bf 1511}, 014 (2015)
  doi:10.1007/JHEP11(2015)014
  [arXiv:1402.6334 [hep-th]].
  
  \bibitem{Maldacena:2016upp} 
  J.~Maldacena, D.~Stanford and Z.~Yang,
  ``Conformal symmetry and its breaking in two dimensional Nearly Anti-de-Sitter space,''
  PTEP {\bf 2016}, no. 12, 12C104 (2016)
  doi:10.1093/ptep/ptw124
  [arXiv:1606.01857 [hep-th]].
  
  \bibitem{Cvetic:2016eiv} 
  M.~Cveti\v{c} and I.~Papadimitriou,
  ``AdS$_{2}$ holographic dictionary,''
  JHEP {\bf 1612}, 008 (2016)
  Erratum: [JHEP {\bf 1701}, 120 (2017)]
  doi:10.1007/JHEP12(2016)008, 10.1007/JHEP01(2017)120
  [arXiv:1608.07018 [hep-th]].
  
  \bibitem{Engelsoy:2016xyb} 
  J.~Engels\"{o}y, T.~G.~Mertens and H.~Verlinde,
  ``An investigation of AdS$_{2}$ backreaction and holography,''
  JHEP {\bf 1607}, 139 (2016)
  doi:10.1007/JHEP07(2016)139
  [arXiv:1606.03438 [hep-th]].
  
  \bibitem{Jensen:2016cah}K.~Jensen,
  ``Chaos in AdS$_2$ Holography,''
  Phys.\ Rev.\ Lett.\  {\bf 117}, no. 11, 111601 (2016)
  doi:10.1103/PhysRevLett.117.111601
  [arXiv:1605.06098 [hep-th]].
  
  \bibitem{Das:2017pif} 
  S.~R.~Das, A.~Jevicki and K.~Suzuki,
  ``Three Dimensional View of the SYK/AdS Duality,''
  JHEP {\bf 1709}, 017 (2017)
  doi:10.1007/JHEP09(2017)017
  [arXiv:1704.07208 [hep-th]].
  
  \bibitem{Das:2017hrt} 
  S.~R.~Das, A.~Ghosh, A.~Jevicki and K.~Suzuki,
  ``Three Dimensional View of Arbitrary $q$ SYK models,''
  JHEP {\bf 1802}, 162 (2018)
  doi:10.1007/JHEP02(2018)162
  [arXiv:1711.09839 [hep-th]].

\bibitem{Das:2017wae} 
  S.~R.~Das, A.~Ghosh, A.~Jevicki and K.~Suzuki,
  ``Space-Time in the SYK Model,''
  JHEP {\bf 1807}, 184 (2018)
  doi:10.1007/JHEP07(2018)184
  [arXiv:1712.02725 [hep-th]].
  
  \bibitem{Yoshida1:2017}
  H.~Kyono, S.~Okumura and K.~Yoshida,
  ``Deformations of the Almheiri-Polchinski model,''
  JHEP {\bf 1703} (2017) 173
  doi:10.1007/JHEP03(2017)173
  [arXiv:1701.06340 [hep-th]].
  
  \bibitem{Lala:2018}
  A.~Lala and D.~Roychowdhury,
  ``SYK/AdS duality with Yang-Baxter deformations,''
  JHEP {\bf 1812} (2018) 073
  doi:10.1007/JHEP12(2018)073
  [arXiv:1808.08380 [hep-th]].
  
  \bibitem{Roychowdhury:2018}
  D.~Roychowdhury,
  ``Holographic derivation of $ q $ SYK spectrum with Yang-Baxter shift,''
  Phys.\ Lett.\ B {\bf 797} (2019) 134818
  doi:10.1016/j.physletb.2019.134818
  [arXiv:1810.09404 [hep-th]].
  
   \bibitem{Davison:2016ngz}
  R.~A.~Davison, W.~Fu, A.~Georges, Y.~Gu, K.~Jensen and S.~Sachdev,
  ``Thermoelectric transport in disordered metals without quasiparticles: The Sachdev-Ye-Kitaev models and holography,''
  Phys.\ Rev.\ B {\bf 95} (2017) no.15,  155131
  doi:10.1103/PhysRevB.95.155131
  [arXiv:1612.00849 [cond-mat.str-el]].
  
 \bibitem{Gaikwad:2018dfc}
A.~Gaikwad, L.~K.~Joshi, G.~Mandal and S.~R.~Wadia,
``Holographic dual to charged SYK from 3D Gravity and Chern-Simons,''
JHEP \textbf{02} (2020), 033
doi:10.1007/JHEP02(2020)033
[arXiv:1802.07746 [hep-th]].
  
  \bibitem{Lala:2019inz}
A.~Lala and D.~Roychowdhury,
``Models of phase stability in Jackiw-Teitelboim gravity,''
Phys. Rev. D \textbf{100} (2019), 124061
doi:10.1103/PhysRevD.100.124061
[arXiv:1909.09828 [hep-th]].
  
  \bibitem{Fan:2015aia} 
  Z.~Y.~Fan and H.~L\"u,
  ``Electrically-Charged Lifshitz Spacetimes, and Hyperscaling Violations,''
  JHEP {\bf 1504}, 139 (2015)
  doi:10.1007/JHEP04(2015)139
  [arXiv:1501.05318 [hep-th]].
  
  \bibitem{Fan:2015yza}
Z.~Y.~Fan and H.~Lu,
``Charged Black Holes in Colored Lifshitz Spacetimes,''
Phys. Lett. B \textbf{743}, 290-294 (2015)
doi:10.1016/j.physletb.2015.02.052
[arXiv:1501.01727 [hep-th]].

\bibitem{Kachru:2008yh}
S.~Kachru, X.~Liu and M.~Mulligan,
``Gravity duals of Lifshitz-like fixed points,''
Phys. Rev. D \textbf{78} (2008), 106005
doi:10.1103/PhysRevD.78.106005
[arXiv:0808.1725 [hep-th]].

\bibitem{Taylor:2008tg} 
  M.~Taylor,
  ``Non-relativistic holography,''
  arXiv:0812.0530 [hep-th].  
  
 \bibitem{Romans:1985ps}
L.~Romans,
``Gauged $N=4$ Supergravities in Five-dimensions and Their Magnetovac
Backgrounds,'' Nucl. Phys. B \textbf{267} (1986), 433-447
doi:10.1016/0550-3213(86)90398-6

\bibitem{Hawking:1983} S~W.~Hawking and D.~N.~Page, ``Thermodynamics
of black holes in anti-de Sitter space," Commun. Math. Phys. 87, 577 (1983).

\bibitem{Chamblin:1999tk}
A.~Chamblin, R.~Emparan, C.~V.~Johnson and R.~C.~Myers,
``Charged AdS black holes and catastrophic holography,''
Phys. Rev. D \textbf{60}, 064018 (1999)
doi:10.1103/PhysRevD.60.064018
[arXiv:hep-th/9902170 [hep-th]].

\bibitem{Chamblin:1999hg}
A.~Chamblin, R.~Emparan, C.~V.~Johnson and R.~C.~Myers,
``Holography, thermodynamics and fluctuations of charged AdS black holes,''
Phys. Rev. D \textbf{60}, 104026 (1999)
doi:10.1103/PhysRevD.60.104026
[arXiv:hep-th/9904197 [hep-th]].

\bibitem{Carlip:2003ne}
S.~Carlip and S.~Vaidya,
``Phase transitions and critical behavior for charged black holes,''
Class. Quant. Grav. \textbf{20} (2003), 3827-3838
doi:10.1088/0264-9381/20/16/319
[arXiv:gr-qc/0306054 [gr-qc]].

\bibitem{Cai:2007wz}
R.~G.~Cai, S.~P.~Kim and B.~Wang,
``Ricci flat black holes and Hawking-Page phase transition in Gauss-Bonnet
gravity and dilaton gravity,''
Phys. Rev. D \textbf{76} (2007), 024011
doi:10.1103/PhysRevD.76.024011
[arXiv:0705.2469 [hep-th]].

\bibitem{Myung:2007ti}
Y.~S.~Myung,
``Phase transition between non-extremal and extremal Reissner-Nordstrom
black holes,''
Mod. Phys. Lett. A \textbf{23} (2008), 667-676
doi:10.1142/S0217732308026741
[arXiv:0710.2568 [gr-qc]].
  
\bibitem{Gibbons:1976ue}
G.~Gibbons and S.~Hawking,
``Action Integrals and Partition Functions in Quantum Gravity,''
Phys. Rev. D \textbf{15}, 2752-2756 (1977)
doi:10.1103/PhysRevD.15.2752

\bibitem{Wald:1993nt}
R.~M.~Wald,
``Black hole entropy is the Noether charge,''
Phys. Rev. D \textbf{48}, no.8, 3427-3431 (1993)
doi:10.1103/PhysRevD.48.R3427
[arXiv:gr-qc/9307038 [gr-qc]].

\bibitem{Jacobson:1993vj}
T.~Jacobson, G.~Kang and R.~C.~Myers,
``On black hole entropy,''
Phys. Rev. D \textbf{49} (1994), 6587-6598
doi:10.1103/PhysRevD.49.6587
[arXiv:gr-qc/9312023 [gr-qc]].

\bibitem{Brustein:2007jj}
R.~Brustein, D.~Gorbonos and M.~Hadad,
``Wald's entropy is equal to a quarter of the horizon area in units of the effective
gravitational coupling,''
Phys. Rev. D \textbf{79} (2009), 044025
doi:10.1103/PhysRevD.79.044025
[arXiv:0712.3206 [hep-th]].

\bibitem{Witten:1998} E.~Witten, ``Anti-de Sitter space, thermal phase transition,
and confinement in gauge theories," Adv. Theor. Math. Phys. 2, 505 (1998).
 


\end{thebibliography}
\end{document}